\documentclass[twocolumn,runningheads]{svjour2}
\smartqed  
\usepackage{graphicx}
\usepackage{amssymb,amsmath,times}
\journalname{Astrophysics and Space Science}
\begin{document}

\title{Numerical studies on the structure of the cosmic ray electron halo in starburst galaxies 
}

\author{Shoko Miyake       \and
        Shohei Yanagita    \and
        Tatsuo Yoshida 
}


\institute{S. Miyake \at
              Faculty of Science, Ibaraki University, Mito 310-8512, Japan \\ 
              JSPS Research Fellow \\
              \email{06nd408g@mcs.ibaraki.ac.jp}           
           \and
           S. Yanagita \at
              Faculty of Science, Ibaraki University, Mito 310-8512, Japan \\
              Tel.: +81-29-228-8394\\
              \email{yanagita@mx.ibaraki.ac.jp}           
           \and
           T. Yoshida \at
              Faculty of Science, Ibaraki University, Mito 310-8512, Japan \\
              \email{yoshidat@mx.ibaraki.ac.jp}           
}

\date{Received: date / Accepted: date}

\maketitle

\begin{abstract}
The structure of the cosmic ray electron halo of a starburst galaxy
depends strongly on the nature of galactic wind and the
configuration of the magnetic field. We have investigated these
dependencies by solving numerically the propagation of electrons 
originating in starburst galaxies, most likely in supernova remnants.
The calculations are made for several models for the
galactic winds and for the configuration of the magnetic fields for
comparison with observations. Our simulation of a quasi-radio halo
reproduces both the extended structure of $\sim 9~\mathrm{kpc}$ and the
subtle hollow structure near the polar region of the radio halo that
are observed in the starburst galaxy NGC\,253. These findings suggest
the existence of strong galactic wind in NGC\,253.

\keywords{cosmic ray halo \and starburst galaxy \and radio halo}
\PACS{95.30.Dr
}
\end{abstract}

\section{Introduction}
\label{intro}

Cosmic ray electrons accelerated in galaxies propagate outward
by diffusion and convection in galactic winds and eventually
form an electron halo, the size of which depends on physical
conditions such as the ambient magnetic field strength and strength of
the galactic wind. The profile of nonthermal radio emission from the
relativistic electrons reveals information on the nature of electron
itself as well as information about the galactic wind from the
starburst region and the configuration of the magnetic field.
We have numerically investigated the structure of the cosmic ray electron
halo of starburst galaxies and the resultant radio
halo to study the existence of the galactic winds and the
configuration of the magnetic field.
The calculation of the propagation of electrons is made by solving a
coupled set of stochastic differential equations (SDE) which is
equivalent to the so-called diffusion convection partial differential
equation and which has been successfully applied to the study of the solar
modulation phenomena of the galactic cosmic rays in the heliosphere.
In this paper we shall use spherically symmetric and axially symmetric
models for the galactic winds and for the configuration of the
magnetic field~\cite{Chevalier,Zirakashvili}.
We present details of the structure of electron halo and also the
resultant structure of radio halo of starburst galaxies, with particular
attention to NGC\,253.

\section{Numerical models}
\label{sec:model}

The SDE equivalent to the diffusion convection partial differential 
equation is written as
\begin{equation}
 \begin{array}{llllll}
d{\bf{X}} & = & {\bf{u}}dt + \sqrt{2\kappa}d{\bf W}(t)~, \\
   & &\\
dP & = & -\frac{1}{3} P\left( \boldsymbol{\nabla} \cdot {\bf{u}} \right)dt 
         -dP_{sync} -dP_{IC}~,
  \end{array}
\label{eq:sde}
\end{equation}
where ${\bf X}$ and $P$ are the position and the momentum of the
particle, ${\bf u}$ is the galactic wind velocity, $P_{sync}$ and
$P_{IC}$ indicate the synchrotron and the inverse Compton momentum
loss, $\kappa$ is the diffusion coefficient, and $d{\bf W}(t)$ is a
Wiener process given by the Gaussian distribution.

We adopted 
\begin{equation}
 \begin{array}{llll}
\kappa & = & 100~{\kappa}_{B} \\
       & = & 3.3 \times 10^{24} \beta \left( \dfrac{P}{1\mathrm{GeV/c}} \right) \left( \dfrac{B}{1\mathrm{\mu G}} \right)^{-1} ~\left[ \mathrm{ cm^2 \cdot sec^{-1}} \right] ,
 \end{array}
\label{eq:kappa}
\end{equation}
\begin{equation}
\dfrac{dP_{sync}}{dt} = \dfrac{4}{3} \sigma_T \beta^{-1}\Gamma^2 \dfrac{B^2}{8\pi}~,
\label{eq:sync}
\end{equation}
\begin{equation}
\dfrac{dP_{IC}}{dt} = \dfrac{4}{3} \sigma_T \beta^{-1}\Gamma^2 U_{ph}~,
\label{eq:IC}
\end{equation}
where ${\kappa}_B$ is the Bohm diffusion coefficient, $B$ is the
magnetic field strength, $\sigma_T$ is the Thomson cross section, 
$\Gamma$ is the Lorentz factor of electrons,
and $U_{ph}$ is the energy density of the 
cosmic microwave background (CMB). 
Here we assume only CMB
photons as the target photons for the inverse Compton process.
In our simulation, particles start at a fixed final point and run
backwards in time until they come to the galactic disk boundary which
has a 7~kpc radius and 0.5~kpc thickness.
The momentum spectrum $f_{\bf{X}}(p)$ at arbitrary position $\bf{X}$
is written as a convolution of the spectrum $f_{\bf{X_0}}(p_0)$ at the
galactic disk boundary with the normalized transition probability
$F(p_0,{\bf{X_0}}|p,{\bf{X}})$ obtained by our SDE method as
\begin{equation}
f_{\bf{X}}(p) = \int f_{\bf{X_0}}(p_0)F(p_0,{\bf{X_0}}|p,{\bf{X}})dp_0~.
\end{equation}
Here the spectrum at the boundary is $f_{\bf{X_0}}(p_0) \propto
(m^2c^4+{p_0}^2c^2)^{-1.6}/p_0$ which is assumed to be uniform at the
galactic disk boundary.
 
The simple analytical model for the spherically symmetric and the
axially symmetric galactic wind flow from a starburst galaxy are given
by Chevalier and Clegg~\cite{Chevalier} and Zirakashvili and
Voelk~\cite{Zirakashvili} respectively. We adopt their models for our
simulations.

For the spherically symmetric model of Chevalier and Clegg, the
analytical solutions in the galactic halo are given as
\begin{equation}
M^{2/(\gamma -1)}\left( \dfrac{\gamma -1+2/M^2}{1+\gamma} \right)^{(\gamma +1)/[2(\gamma -1)]} = \left( \dfrac{r}{R} \right)^2~,
\label{eq:1D-1}
\end{equation}
\begin{equation}
\rho ur^2=\mathrm{const}~,
\label{eq:1D-2}
\end{equation}
\begin{equation}
\rho ur^2\left( \dfrac{1}{2}u^2+\dfrac{\gamma}{\gamma -1}
\dfrac{P}{\rho} \right) = \mathrm{const}~,
\label{eq:1D-3}
\end{equation}
where $M$ is the Mach number, $r$ is the radial coordinate, $R$ is
radius of the wind base, $\rho$ is the density, $P$ is the pressure,
$u$ is the galactic wind velocity, and $\gamma$ is the adiabatic
index. We adopted $R=300~\mathrm{pc}$, $\gamma = 5/3$, the mass loss
rate ${\dot M} = 3.9 M_\odot~\mathrm{yr^{-1}}$, and the energy
production rate ${\dot E} = 1.9 \times 10^{42}~\mathrm{erg \cdot
  s^{-1}}$. We also assume the magnetic field is frozen in the
galactic wind flow, namely $B(r) = \left( \rho (r)/ \rho (R)
\right)^{2/3}B_0$, where $B_0$ is the $B$ at $r=R$. $B_0$ is assumed
to be $50~\mathrm{\mu G}$.

For the axially symmetric model of Zirakashvili and Voelk, the analytical 
solutions in the galactic halo are given as 
\begin{equation}
u_r = u_\infty \mathrm{cos}\theta~,
\label{eq:2D-Vr}
\end{equation}
\begin{equation}
u_\theta = -\left( 1-\gamma^{-1} \right) u_\infty \mathrm{sin}\theta~,
\label{eq:2D-Vt}
\end{equation}
\begin{equation}
\rho (r,\theta) = \rho_g \mathrm{sin^{-2\frac{\gamma -1}{2\gamma -1}}}\theta
 \left( \dfrac{r}{R_g} \right)^{-\frac{2\gamma}{2\gamma -1}}~,
\label{eq:2D-rho}
\end{equation}
\begin{equation}
p(r,\theta) = \rho_g \mathrm{sin^{\frac{2\gamma}{2\gamma -1}}}\theta
 \left( \dfrac{r}{R_g} \right)^{-\frac{2\gamma}{2\gamma -1}}~,
\label{eq:2D-p}
\end{equation}
where $u_r$ and $u_\theta$ are the radial and latitudinal velocity
component in cylindrical coordinates $(r,\theta,z)$, $\rho$ is the
density, $p$ is the pressure, $u_\infty$ is the asymptotic velocity of
the galactic wind, $\gamma$ is the adiabatic index, $R_g$ is the
radius of the wind base, and $\rho_g$ is the $\rho$ at $r=R_g$. We
adopted $\gamma = 5/3$, $R_g = 300~\mathrm{pc}$ and $u_\infty =
900~\mathrm{km/sec}$.

We used the following expression~\cite{Zirakashvili} for the magnetic field which has a 
similar $r$-dependence as $p(r,\theta)$,
\begin{equation}
B=B_0\left(\mathrm{sin}\theta \right)^{\frac{1}{\gamma -1}\left(\gamma_m
  -\frac{\gamma^2}{2\gamma -1} \right)}\left( \dfrac{r}{R_g} \right)^{-
  \frac{\gamma}{2\gamma -1}}~,
\label{eq:2D-b}
\end{equation}
where $B_0$ is the magnetic field strength at the wind base 
and $\gamma_m$ is the adiabatic index of an isotropic random 
magnetic field. We assumed $B_0 = 50~\mathrm{\mu G}$ and $\gamma_m = 4/3$.

\section{Results}
\label{sec:res}
\subsection{Structure of cosmic ray electron halo}
\label{subsec:CR_halo}

\begin{figure}[tbh]
\centering
  \begin{minipage}{0.48\hsize}
  \begin{center}
    \includegraphics[scale  =0.21, bb=88 104 626 663,clip]{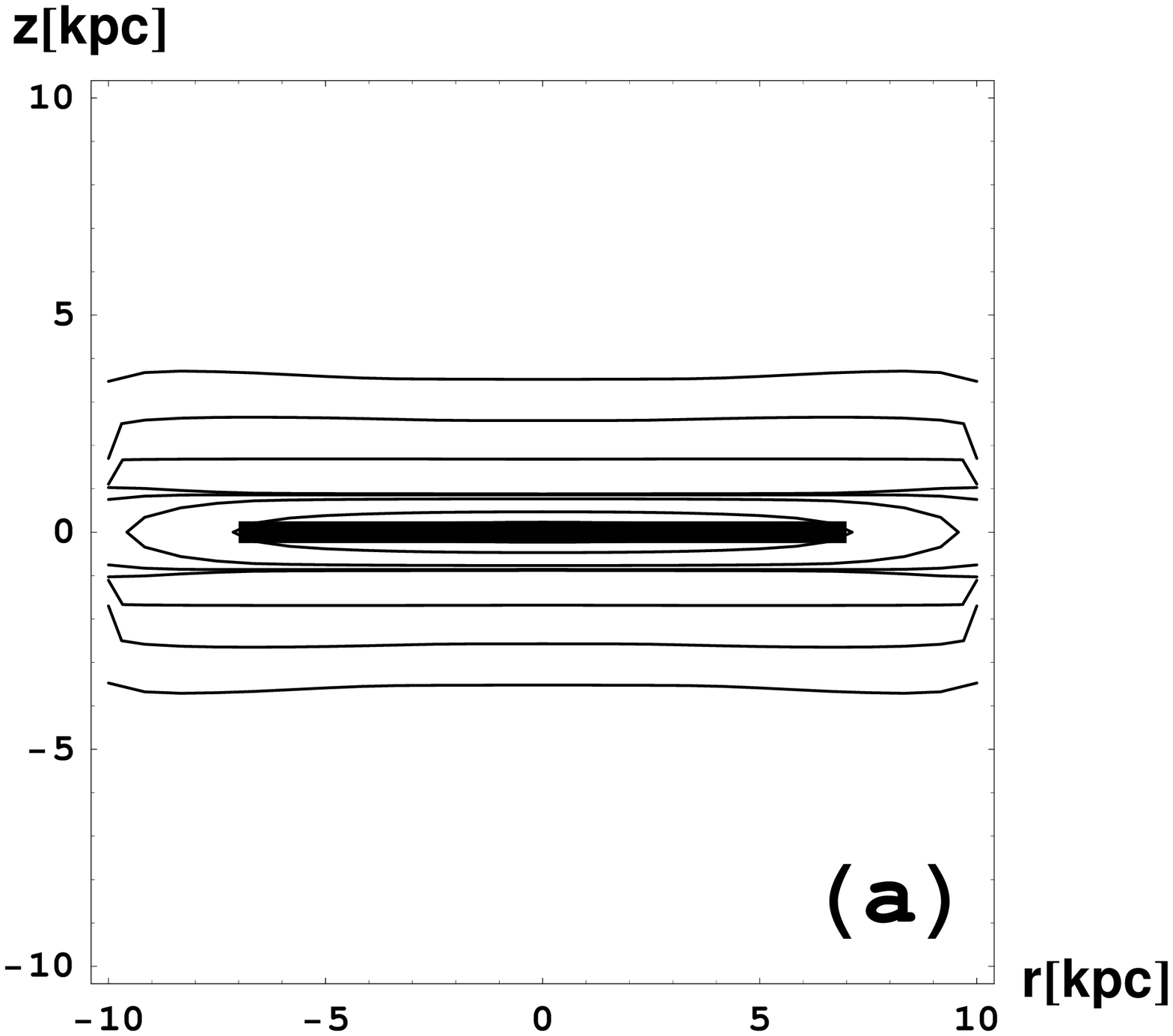}
  \end{center}
  \end{minipage}
  \begin{minipage}{0.48\hsize}
  \begin{center}
    \includegraphics[scale =  0.21, bb=88 104 626 663,clip]{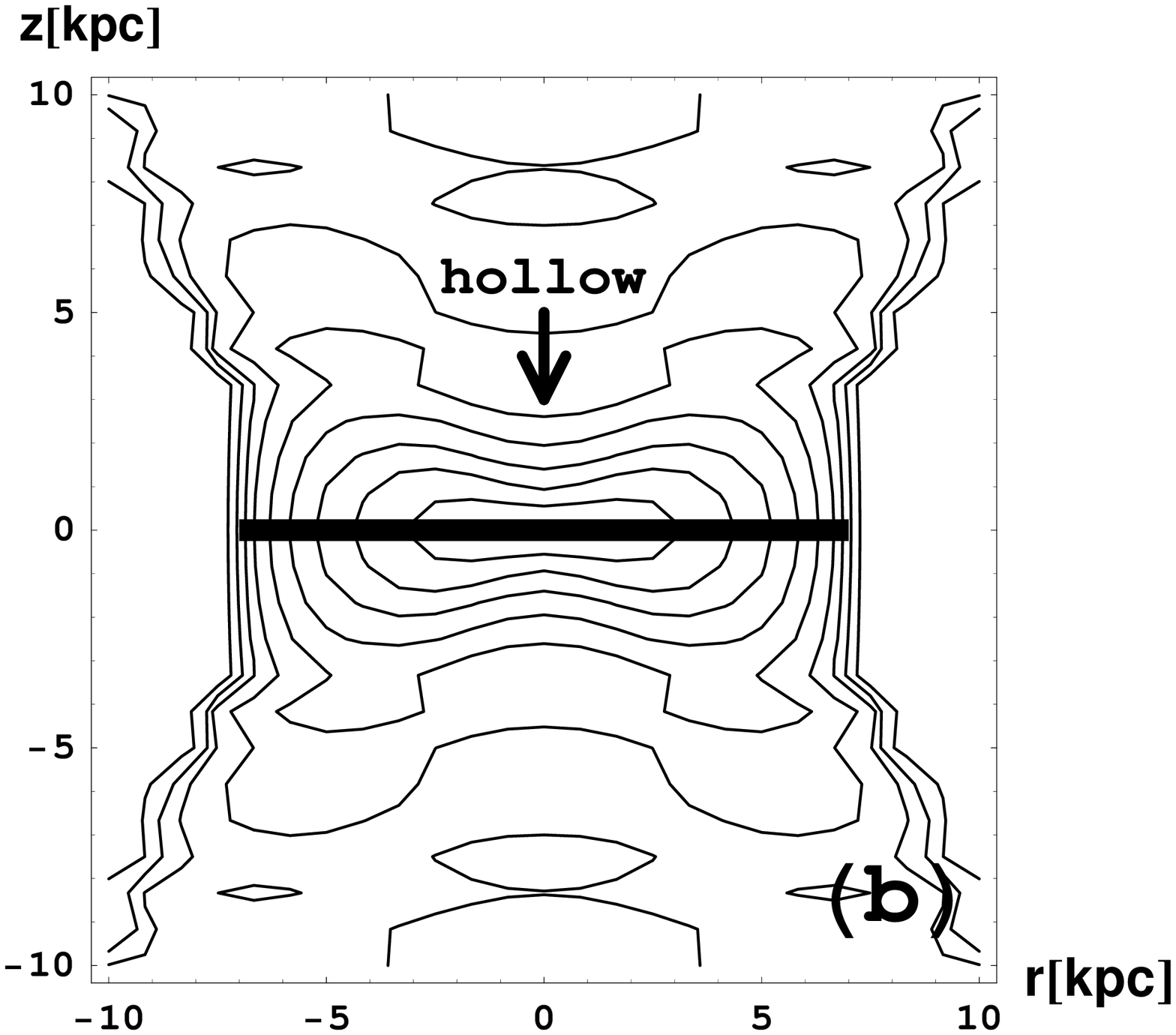}
  \end{center}
  \end{minipage}
\caption{(a) Side view of 1~GeV electron intensities integrated along the line of sight in the spherically symmetric model. The black rectangle indicates the galactic disk boundary. The peak intensity is normalized to 1.0. Contours indicate $10^{-6}$, $10^{-5}$, $10^{-4}$, $10^{-3}$, 0.01, 0.1, 0.5, and 0.8.; (b) Same as (a) in the axially symmetric model. Contours indicate 0.1, 0.2, 0.3,..., 0.8, and 0.9.
}
\label{fig:md}    
\end{figure}

\begin{figure}[t]
\centering
    \includegraphics[scale =  0.3, bb=88 104 626 663,clip]{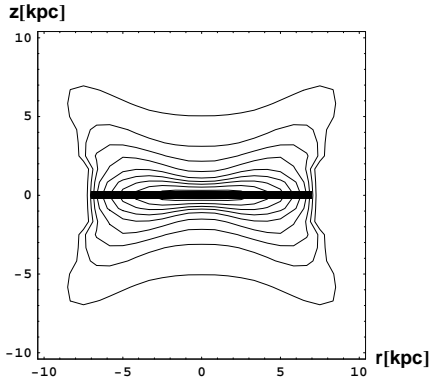}
\caption{Side view of electron intensities integrated along the line of sight in the axially symmetric model for 10~GeV electrons. The black rectangle indicates the galactic disk boundary. The peak intensity is normalized to 1.0. Contours indicate 0.1, 0.2, 0.3,..., 0.8, and 0.9.}
\label{fig:ed}
\end{figure}

\begin{figure}[t]
\centering
    \includegraphics[scale =  0.3, bb=88 104 626 663,clip]{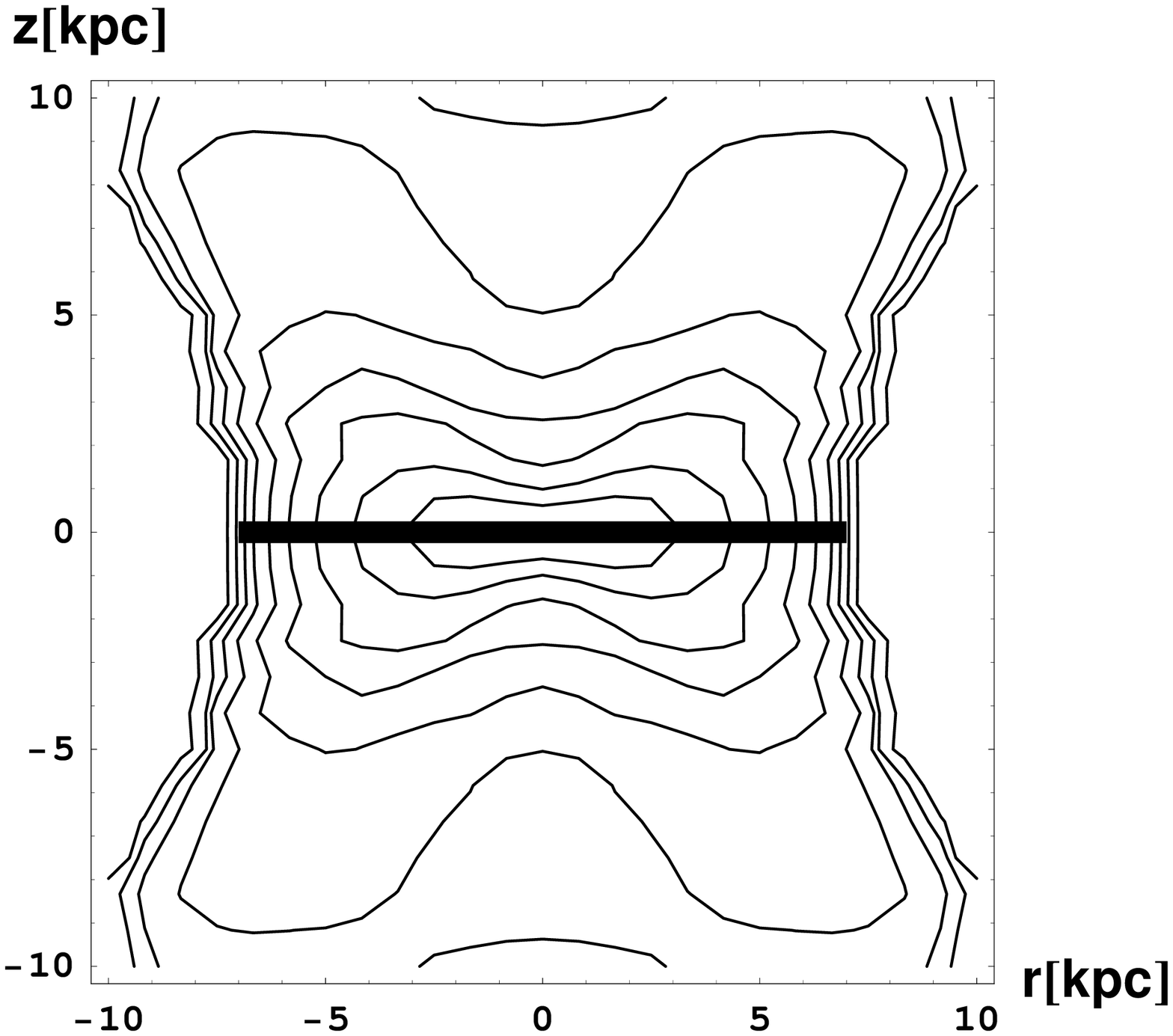}
\caption{Side view of 1~GeV electron intensities integrated along the line of sight in the axially symmetric model for a $B_0$ of 5~$\mathrm{\mu G}$. The black rectangle indicates the galactic disk boundary. The peak intensity is normalized to 1.0. Contours indicate 0.1, 0.2, 0.3,..., 0.8, and 0.9.}
\label{fig:bd}  
\end{figure}

Fig.~\ref{fig:md} shows side views of the 1~GeV electron intensities
integrated along the line of sight in the spherically symmetric and
the axially symmetric models. The filled rectangle at the center indicates
the side view of the galactic disk.
We can see in Fig.~\ref{fig:md} there is a big difference in the
structure and the size of the resultant electron halo of the two
models for symmetry. One reason for this difference comes from the
difference in the distribution of the source position of electrons on
the galactic disk boundary. In the axially symmetric model electrons may
come from almost any point on the galactic disk boundary, however, in
spherical symmetric model almost all electrons should originate near
the galactic center where the magnetic field intensity is high and
electrons suffer severe energy loss. The spherical symmetry model clearly
does not reproduce the observed radio halo of NGC\,253 \cite{Carilli}.
We discuss below only the results for axially symmetric models,
because in many actual cases the starburst region has an ellipsoidal
shape or a disk geometry \cite{Zirakashvili}.

Propagation of electrons is governed by the diffusion process and 
convection in the expanding galactic wind and the associated energy loss
processes. The diffusion process depends on the electron energy,
however the convection does not.  The adiabatic loss is proportional
to the electron energy, and the other energy loss processes are
proportional to the square of electron energy. The synchrotron
energy loss is larger than the inverse Compton energy loss, because
$B^2/8\pi \gg U_{ph}$. Accordingly the resultant structure depends on
the energy of the electrons as we will see below.  Fig.~\ref{fig:ed} shows
a side view of the electron intensities integrated along the line of sight
in the axially symmetric model for 10~GeV electrons. The contours
in Fig.~\ref{fig:ed} were drawn with the same steps in
Fig.~\ref{fig:md}.b for easier comparison.  The higher the electron
energy, the smaller the size of the halo, due to the shorter
synchrotron cooling time.  For the axially symmetric model the TeV
electron halo is not formed, because the synchrotron cooling time of
TeV electrons is ${\sim}~5\times 10^3$~years, and the diffusion length
is only about 1~pc.  This is consistent with the H.E.S.S.\ result of
TeV $\mathrm{\gamma}$-ray observation of NGC 253~\cite{HESS}.  The
synchrotron energy loss is proportional to the square of magnetic
field strength, therefore the weaker the magnetic field strength,
the larger the size of the halo.  Fig.~\ref{fig:bd} shows a side
view of the 1~GeV electron intensities integrated along the line of
sight in the axially symmetric model for a $B_0$ of $5~\mathrm{\mu G}$.
The size of the halo shown in Fig.~\ref{fig:bd} is much larger than
that shown in Fig.\ref{fig:md}b as expected.

\begin{figure}[tbh]
\centering
    \includegraphics[scale =  0.45, bb=88 294 626 473,clip]{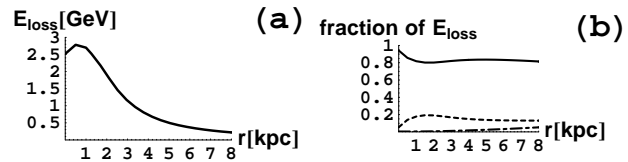}
\caption{(a) Mean energy loss of electrons arriving at $z = 5$\,kpc with an energy of 1~GeV as a function of $r$. (b) The fraction of energy loss for three processes corresponding to the mean energy loss shown in (a). The solid line indicates the adiabatic energy loss. The dotted line indicates the synchrotron radiation loss. The dot-dash line indicates the inverse Compton radiation loss.
}
\label{fig:el}  
\end{figure}

A slight hollow structure near the polar region in
electron halo is visible in Fig.~\ref{fig:md}b and Fig.~\ref{fig:bd} for 
1~GeV electrons and in Fig.~\ref{fig:ed} for 10~GeV electrons. This
structure comes from the $r$ dependence of energy loss in cylindrical
coordinates $(r, \theta, z)$.
Fig.~\ref{fig:el}a shows the mean total energy loss of 1~GeV electrons
at $z = 5$\,kpc as a function of $r$. The energy
loss peaks near the polar region, as the energy loss
is dominated by the adiabatic loss, as shown in Fig.~\ref{fig:el}b,
by the convection due to the galactic wind and as the wind velocity is
higher in polar region as shown by Eq.\ref{eq:2D-Vr}.
Fig.~\ref{fig:el}b shows the $r$ dependence of the fraction of the
energy loss shown in Fig.~\ref{fig:el}a for the three processes,
adiabatic, synchrotron, and inverse Compton.  The slight hollow
structure for 10~GeV electrons seen in Fig.~\ref{fig:ed} comes from
the dominance of synchrotron loss over the other two energy loss
mechanisms. (Due to page limitations we cannot present a similar figure to
Fig.~\ref{fig:el} in this paper.) The rate of synchrotron
energy loss is proportional to the square both of energy and
magnetic field strength. The magnetic field strength is higher in the
polar region as shown by Eq.\ref{eq:2D-b}. Electrons which arrive at
the polar region originate from near the galactic center where they
suffer severe energy loss at the start of their journey.

\begin{figure}[t]
\centering
    \includegraphics[scale =  0.3]{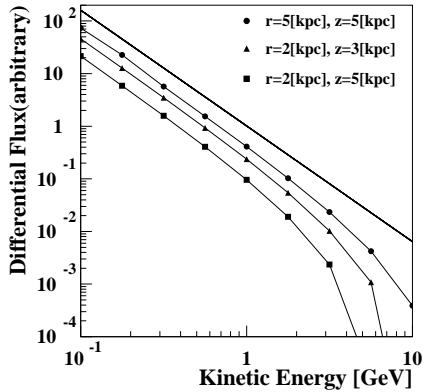}
\caption{Simulated electron energy spectra at several selected positions. The solid lines indicate the energy spectra in the electron halo. The dotted line indicates the energy spectrum in the galactic disk.}
\label{fig:spec}
\end{figure}

The differential electron energy spectrum varies with position
due to the variation of the total energy loss.  Fig.~\ref{fig:spec}
shows the simulated electron energy spectra at several selected
positions. The modulated spectrum shifts from the spectrum at the
galactic disk boundary to the low energy region without changing shape
while the energy loss is dominated by the adiabatic loss, namely
$d{\mathrm{log}}E \propto {\mathrm{const}}$, where $E$ is the electron energy.
But the spectrum steepens in the high energy region
due to the synchrotron energy loss.  The bending point of the spectrum
appears as expected at the electron energy where the adiabatic loss
rate equals to the synchrotron energy loss rate~\cite{Lerche}.

\subsection{Comparison with observation}
\label{subsec:obs}

NGC 253 is a nearby edge-on starburst galaxy at a distance of 2.5~Mpc. The
galaxy has two spiral arms, and a bar of projected length 7~kpc. It
has an extended nonthermal radio halo of $\sim$9~kpc. It has also
extended X-ray halo, possibly related to the galactic wind.

\begin{figure}[tbh]
\centering
    \includegraphics[scale =  0.3, bb=88 104 626 663,clip]{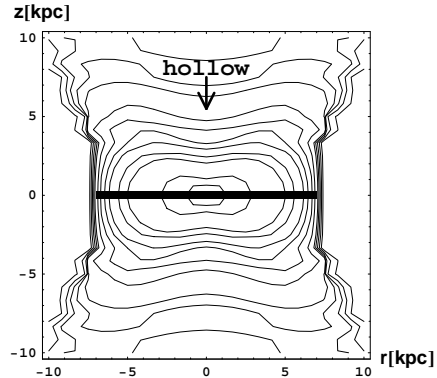}
\caption{Side view of electron (1~GeV) intensities times $B^2$ integrated along the line of sight in the axially symmetric model. The black rectangle indicates the galactic disk boundary. The peak value is normalized to 1.0. Contours indicate 0.005, 0.01, 0.015, 0.02, 0.025, 0.0375, 0.05, 0.0625, 0.075, 0.1, 0.125, 0.25, and 0.5.
}
\label{fig:simu} 
\end{figure}

Fig.~\ref{fig:simu} shows side view of the 1~GeV electron intensities
multiplied by $B^2$ and integrated along the line of sight in the axially
symmetric model. As the power of synchrotron radiation is proportional to
electron intensity times the square of magnetic field strength,
Fig.~\ref{fig:simu} may mimic the radio intensity profile.

Our result for the quasi-radio intensity profile qualitatively reproduces 
the observed large radio halo of $\sim$~9~kpc. The contours in
Fig.~\ref{fig:simu} are drawn with steps proportional to the
steps in Fig.~2 of Carilli et al.\ \cite{Carilli} for
ease of comparison of the results of our simulations with the observed radio
halo. Looking closer at the observed halo, shown in Fig.~2 of Carilli
et al.\ \cite{Carilli}, we can recognize a slight hollow
near the polar region and prominence far above the disk. Our
quasi-radio halo reproduce these two features as seen in
Fig.~\ref{fig:simu}. The hollow structure comes from adiabatic cooling
of electrons by the higher velocity galactic wind near the polar
region as shown in Fig.~\ref{fig:md}b. Our results suggest the
existence of a strong galactic wind in NGC~253. In our simulation, the
cosmic ray sources are assumed to be distributed uniformly in the
galactic disk. The nature of galactic winds and magnetic fields in
starburst galaxies will be revealed by further simulation experiments
based on models which take into account a more realistic
distribution of cosmic ray sources in galaxies.

\section{Conclusions}
\label{sec:conc}

We have examined numerically the structure of the cosmic ray electron halo
of starburst galaxies.  We confirmed that the formation of electron
halo extending to $\sim 9~\mathrm{kpc}$ from galaxies is possible 
for the axially
symmetric model of the galactic wind flow and the magnetic field
configuration.  Furthermore, a subtle hollow structure of radio halo
near the polar region observed in NGC\,253 was reproduced.  This
structure may come from large energy loss near at the polar region by
either the adiabatic energy loss due to strong galactic wind or the
synchrotron energy loss due to relatively strong magnetic field there.
These findings suggest the existence of strong galactic wind in NGC\,253.

\begin{acknowledgements}
We are grateful to Phil Edwards for careful reading of the manuscript
and important suggestions.
S. Miyake is supported by a JSPS Research Fellowship.
\end{acknowledgements}

\end{document}